\documentclass[a4paper,conference]{IEEEtran}
\usepackage{url}
\usepackage{hyperref}
\usepackage{graphicx}
\usepackage{amsmath,graphicx}
\usepackage{amssymb}
\usepackage{pifont}
\newcommand{\cmark}{\ding{51}}
\newcommand{\xmark}{\ding{55}}
\usepackage{xcolor}

\ifCLASSINFOpdf
\else
\fi

\title{Automatic Foot Ulcer Segmentation Using an Ensemble of Convolutional Neural Networks\thanks{© 2022 IEEE. Personal use of this material is permitted. Permission from IEEE must be obtained for all other uses, in any current or future media, including
reprinting/republishing this material for advertising or promotional purposes, creating new collective works, for resale or redistribution to servers or lists, or reuse of any copyrighted component of this work in other works.
}
}

\author{
	\IEEEauthorblockN{Amirreza Mahbod} 
	\IEEEauthorblockA{\textit{Institute for Pathophysiology and Allergy Research}\\
		\textit{Medical University of Vienna}\\
		Vienna, Austria}\\  
	\IEEEauthorblockN{Rupert Ecker} 
	\IEEEauthorblockA{\textit{Research and Development Department}\\
		\textit{TissueGnostics GmbH}\\
		Vienna, Austria}
	\and
	\IEEEauthorblockN{Gerald Schaefer} 
	\IEEEauthorblockA{\textit{Department of Computer Science}\\
		\textit{Loughborough University}\\
		Loughborough, United Kingdom}\\ 
	\IEEEauthorblockN{Isabella Ellinger} 
	\IEEEauthorblockA{\textit{Institute for Pathophysiology and Allergy Research}\\
		\textit{Medical University of Vienna}\\
		Vienna, Austria}

}
\IEEEoverridecommandlockouts
\begin{document}
	\maketitle

\begin{abstract}
Foot ulcer is a common complication of diabetes mellitus and, associated with substantial morbidity and mortality, remains a major risk factor for lower leg amputations. Extracting accurate morphological features from foot wounds is crucial for appropriate treatment. Although visual inspection by a medical professional is the common approach for diagnosis, this is subjective and error-prone, and computer-aided approaches thus provide an interesting alternative. Deep learning-based methods, and in particular convolutional neural networks (CNNs), have shown excellent performance for various tasks in medical image analysis including medical image segmentation.

In this paper, we propose an ensemble approach based on two encoder-decoder-based CNN models, namely LinkNet and U-Net, to perform foot ulcer segmentation. To deal with a limited number of available training samples, we use pre-trained weights (EfficientNetB1 for the LinkNet model and EfficientNetB2 for the U-Net model) and perform further pre-training using the Medetec dataset while also applying a number of morphological-based and colour-based augmentation techniques.  To boost the segmentation performance, we incorporate five-fold cross-validation, test time augmentation and result fusion.

Applied on the publicly available chronic wound dataset and the MICCAI 2021 Foot Ulcer Segmentation (FUSeg) Challenge, our method achieves state-of-the-art performance with data-based Dice scores of 92.07\% and 88.80\%, respectively, and is the top ranked method in the FUSeg challenge leaderboard. The Dockerised guidelines, inference codes and saved trained models are publicly available at \url{https://github.com/masih4/Foot_Ulcer_Segmentation}.  

\end{abstract}

\begin{IEEEkeywords}
	Foot ulcer, medical image analysis, image segmentation, machine learning, deep learning, ensemble.
\end{IEEEkeywords}

\IEEEpeerreviewmaketitle

\section{Introduction}
\label{sec:intro}
Diabetes mellitus is one of the most common chronic diseases with an increasing rate of prevalence over the past decades~\cite{GUARIGUATA2014137}. It can cause several complications for the patient such as cardiovascular disease, retinopathy and neuropathy~\cite{harding2019global}. A serious medical condition associated with diabetes mellitus are skin ulcers on the foot, which can lead to amputation~\cite{BANDYK201843}. Up to 3\% of patients with diabetes mellitus have an active foot ulcer, while lifetime risk of developing a foot ulcer is up to 25\%. Different treatments may be applied based on the ulcer type and its morphological appearance~\cite{doi:10.1177/0141076816688346}.

For appropriate treatment, lesion characteristics such as length, width, area, and volume need to be measured. Moreover, the wound area is a useful predictor of the final outcome and allows to monitor the healing process and thus to evaluate the effect of treatment~\cite{https://doi.org/10.1111/iwj.12472}. Visual inspection and measurement of foot ulcers by a medical expert is the common approach to investigate these morphological features. However, this is subjective and error-prone~\cite{niri2021superpixel}. A promising alternative is to acquire images of the area and apply computer-aided methods to segment the ulcer. Following automated wound segmentation, the morphological features such as lesion width or area can then be easily extracted from the segmentation masks using morphological operations.


In the literature, a number of computerised methods have been proposed to perform foot ulcer segmentation. These include approaches ranging from classical computer vision techniques to state-of-the-art machine learning and deep learning models. Clustering, edge detection, adaptive thresholding, and region growing are examples of conventional image analysis approaches for wound segmentation~\cite{6392633, AHMADFAUZI201574}, while, based on hand-crafted features, classifiers such as multi-layer perceptrons and support vector machine can also be trained for foot ulcer segmentation~\cite{6392633, 7755785}. Similar to other medical image segmentation tasks,
deep learning and convolutional neural network (CNN)-based approaches have been shown to outperform other approaches for foot ulcer segmentation~\cite{wang2020fully}. CNN-based architectures employed for this task include fully convolutional neural networks (FCNs), U-Net, and Mask R-CNN~\cite{wang2020fully,munoz2020automatic, 10.1007/978-3-030-51935-3_17}.    

\begin{figure*}[t!]
	\centering
	\includegraphics[width=0.7\linewidth]{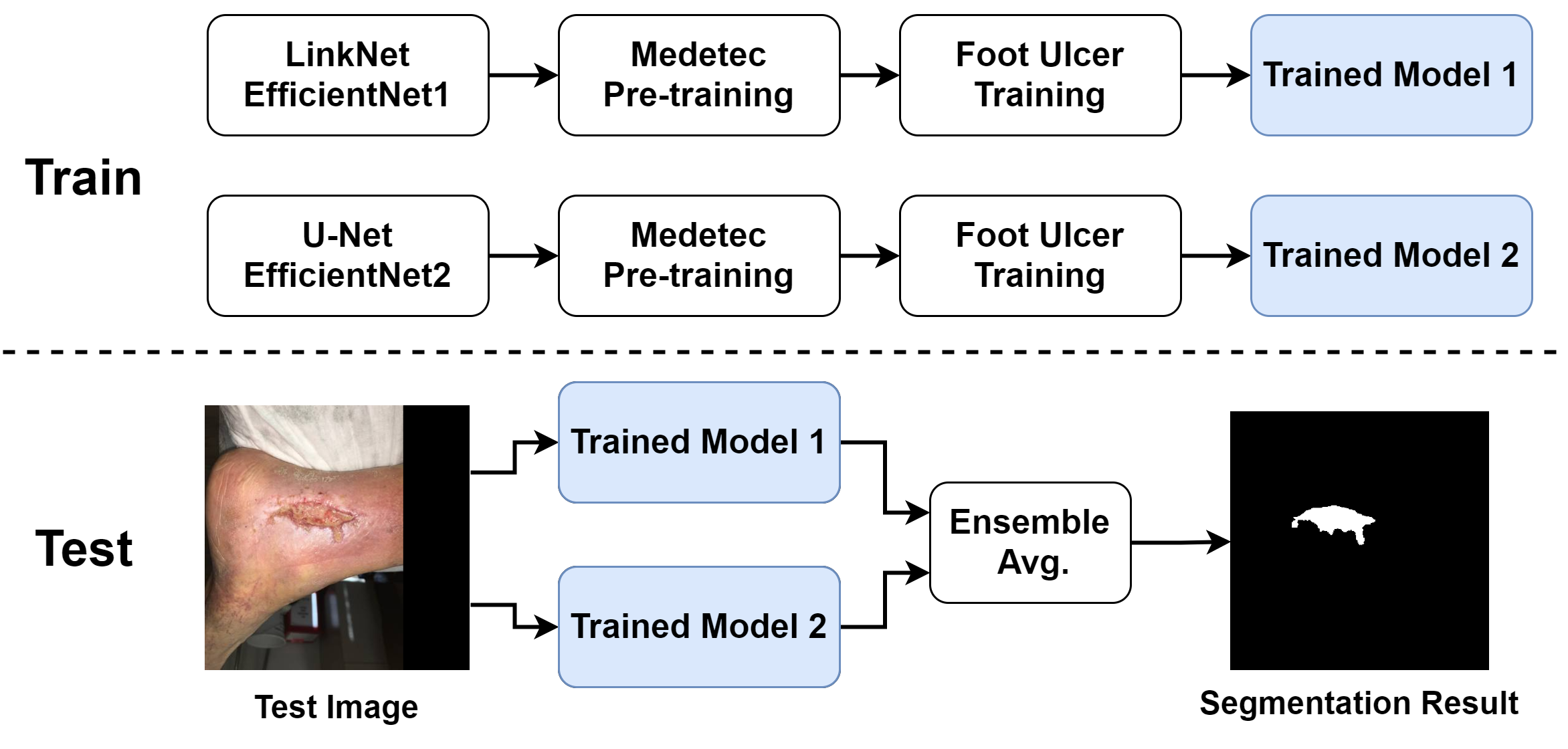}
	\caption{Workflow of the proposed method.}
	\label{flowchart}
\end{figure*}

In this paper, inspired by our former work on other medical image segmentation tasks~\cite{MAHBOD2020105725, mahbod2021cryonuseg_org}, we propose and develop a model based on two well-established encoder-decoder-based CNNs, namely U-Net and LinkNet, to segment wounds in clinical foot images.  We use pre-trained CNNs, 5-fold cross-validation, test time augmentation (TTA), and result fusion to boost the segmentation performance. We evaluate our method on the publicly available chronic wound dataset~\cite{wang2020fully} and the MICCAI 2021 Foot Ulcer Segmentation (FUSeg) Challenge dataset and achieve state-of-the-art segmentation performance for both cases while achieving first rank for the latter\footnote{Challenge leaderboard (accessed on 2022-05-18): \url{https://uwm-bigdata.github.io/wound-segmentation/}}.

\section{Method}
\label{sec:majhead}

The generic workflow of our proposed method, for both training and testing phases, is shown in Fig.~\ref{flowchart}.

\subsection{Datasets}
\label{Datasets}
\begin{figure*}[t!]
	\centering
	\includegraphics[width=0.65\linewidth]{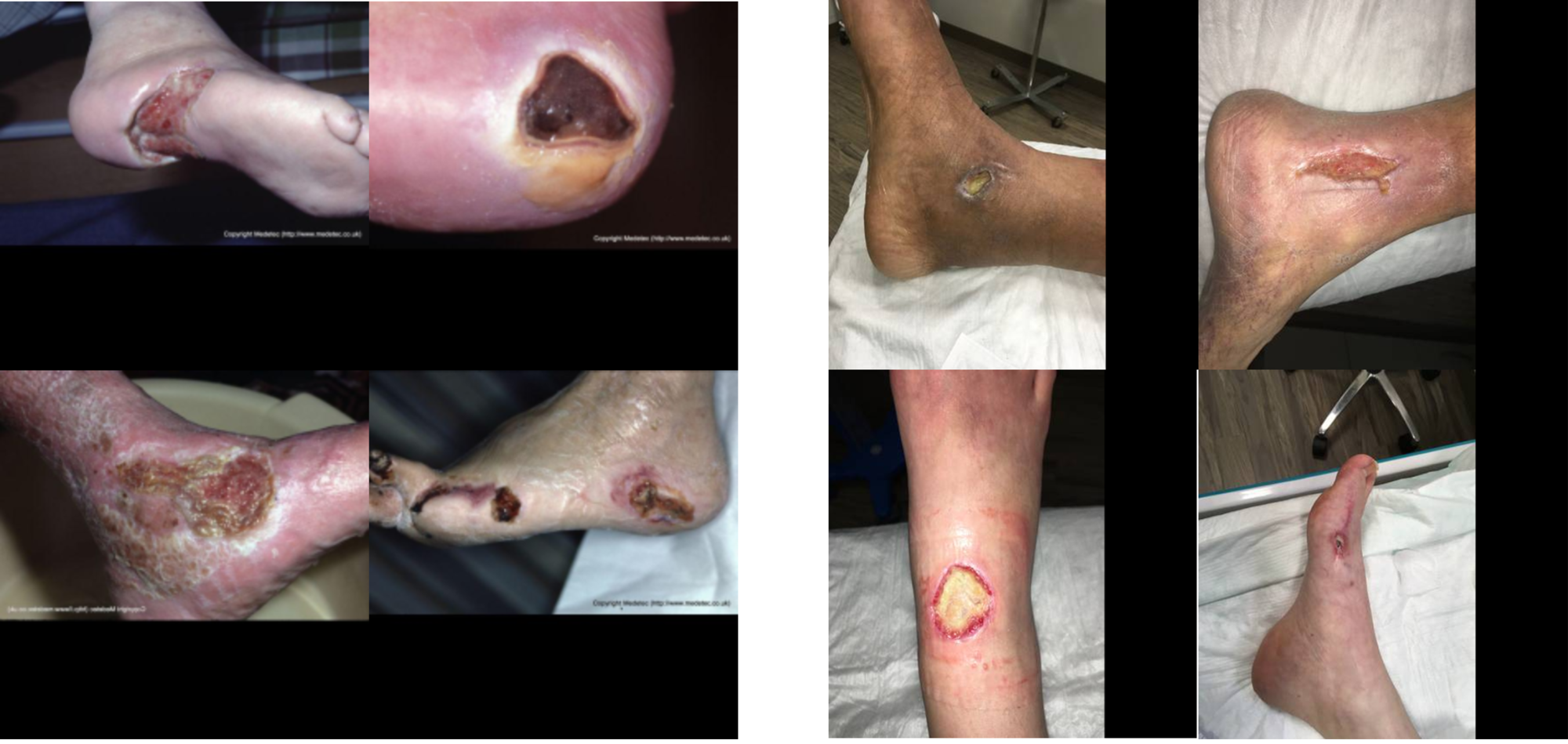}
	\caption{Image samples from the Medetec dataset (left) and the chronic wound dataset/FuSeg dataset (right).}
	\label{samples}
\end{figure*}

We use the following publicly available datasets:
\begin{itemize}
\item
\textbf{Medetec dataset}~\cite{medetec}: consists of 152 clinical images, with corresponding segmentation masks, of several types of open foot wounds. The images are zero-padded and have a fixed size of $224 \times 224$ pixels~\cite{wang2020fully}. We use this dataset only for pre-training of the employed segmentation models.
\item
\textbf{Chronic wound dataset}~\cite{wang2020fully}: contains 1010 clinical images from 889 patients, with the corresponding segmentation masks of all images also available. To obtain a fixed size for all images, a YOLOV3 object detection model~\cite{redmon2018yolov3} is used to localise and crop the wounds inside the images. For non-square images, zero-padding is used to yield a fixed size of $512 \times 512$ pixels for the entire dataset. Of the 1010 images, 810 images are used for training and the remaining 200 kept aside for testing as suggested in~\cite{wang2020fully}.    
\item
\textbf{FUSeg dataset}~\cite{wang2022fuseg}: this dataset is an extended version of the chronic wound dataset and is the dataset used as the training and testing sets of the MICCAI 2021 Foot Ulcer Segmentation Challenge. It contains 1210 images where 1010 images (identical to the chronic wound dataset) are provided for training and 200 images are used for evaluation. The ground truth segmentation masks of the 200 test images are kept private by the challenge organisers and are used only for evaluation purposes. Similar to the chronic wound dataset, all images have a fixed size of $512 \times 512$ pixels. Further details about this dataset can be found on the challenge platform\footnote{\url{https://fusc.grand-challenge.org/}} and in~\cite{wang2022fuseg}. 
\end{itemize}
Fig.~\ref{samples} shows some example images from the datasets.

\subsection{CNN models}
\label{CNN models}
In our approach, we employ two CNN models, namely U-Net~\cite{Ronneberger2015} and LinkNet~\cite{8305148}. Although other encoder-decoder-based CNNs such as U-Net++~\cite{10.1007/978-3-030-00889-5_1} or attention U-Net~\cite{oktay2018attention} can also be used, the chosen CNN models have shown excellent segmentation performance for a variety of medical image segmentation tasks while being computability less expensive~\cite{MAHBOD2020105725, mahbod2021cryonuseg_org, 9446924}. For both models, we use pre-trained CNNs in the decoder part of the model, for LinkNet a pre-trained EfficientNetB1~\cite{tan2019efficientnet}, and for U-Net a pre-trained EfficientNetB2~\cite{tan2019efficientnet}. EfficientNet-based models are state-of-the-art classification models employed in various medical image classification tasks such as skin lesion classification or diabetic retinopathy detection~\cite{MAHBOD2020105475, tymchenko2020deep}. We therefore employ them as the backbone of our encoder-decoder-based segmentation models. In particular, we use two variants of the EfficientNet model, B1 and B2, since one is originally designed for smaller images (and hence more suitable for smaller lesions), while the other is designed for slightly larger images (and hence more suitable for larger lesions)~\cite{tan2019efficientnet}.

\subsection{Training}
As shown in Fig.~\ref{flowchart}, we use the Medetec dataset as a task-specific dataset for further pre-training. During training, we use random scaling (scale limit 0.1 with 0.3 probability), random 90-degree rotations (with 0.5 probability), vertical and horizontal flipping (with 0.5 probability), as well as brightness and contrast shifts (limit of 0.15 with 0.4 probability) as augmentation techniques. We use the full-size images and train each model for 80 epochs with a learning rate (LR) scheduler that reduces the LR by 90\% after every 25 epochs, and an initial LR of 0.001. We use a batch size of 4 and employ the Adam optimiser~\cite{Kingma2014}. As loss function, we use a combination of Dice loss and focal loss (with equal weights). For each dataset, five-fold cross-validation is exploited, and the best models based on segmentation scores on the validation sets are saved to be used in the inference phase. 

\begin{table*}[t!]
	\caption{Segmentation results on the chronic wound data set obtained by methods reported in~\cite{wang2020fully} (top) in comparison to our approach (bottom).}
	\centering
	\label{wound_res}
	\begin{tabular}{lccccc}
		\hline \hline
		&\textbf{\begin{tabular}[c]{@{}l@{}}image-based\\ Dice [\%]\end{tabular}}& \textbf{precision [\%]} & \textbf{recall [\%]}   & \textbf{\begin{tabular}[c]{@{}l@{}}data-based\\ IoU [\%]\end{tabular}} & \textbf{\begin{tabular}[c]{@{}l@{}}data-based\\ Dice [\%]\end{tabular}}\\
		\hline \hline
		VGG16             &n/a  &83.91  &78.35  &n/a  &81.03\\
		SegNet            &n/a  &83.66  &86.49  &n/a  &85.05\\
		U-Net             &n/a  &89.04  &91.29  &n/a  &90.15 \\
		Mask-RCNN         &n/a  &\textbf{94.30}  &86.40  &n/a  &90.20 \\
		MobileNetV2       &n/a  &90.86  &89.76  &n/a  &90.30 \\
		MobileNetV2 + CCL  &n/a  &91.01  &89.97  &n/a  &90.47 \\
		\hline
		LinkNet-EffB1    &83.93   &92.88   &91.33  &85.35   &\textbf{92.09}\\
		U-Net-EffB2       &84.09   &92.23   &91.57  &85.01   &91.90\\
		Ensemble           &\textbf{84.42}   &92.68   &\textbf{91.80}  &\textbf{85.51}   &92.07\\
		\hline \hline
	\end{tabular}
\end{table*}

\begin{table*}
	\caption{Results of ablation study to show the effectiveness of five-fold cross-validation (CV) and test time augmentation (TTA) on the segmentation scores. The best result for each evaluation measure for each of the models is bolded.}
	\centering
	\label{ablation}
	\begin{tabular}{lccccccc}
		\hline \hline
		&\textbf{CV} & \textbf{TTA} &\textbf{\begin{tabular}[c]{@{}l@{}}image-based\\ Dice [\%]\end{tabular}}& \textbf{precision [\%]} & \textbf{recall [\%]}   & \textbf{\begin{tabular}[c]{@{}l@{}}data-based\\ IoU [\%]\end{tabular}} & \textbf{\begin{tabular}[c]{@{}l@{}}data-based\\ Dice [\%]\end{tabular}}\\
		\hline \hline
		LinkNet-EffB1   &\xmark   & \xmark    &82.98   &91.92   &90.39  &83.37   &91.15\\
		LinkNet-EffB1   &\xmark   & \cmark    &82.98   &92.38   &90.83  &84.50   &91.60\\
		LinkNet-EffB1   &\cmark   & \cmark    &\textbf{83.93}   &\textbf{92.88}   &\textbf{91.33}  &\textbf{85.35}   &\textbf{92.09}\\
		\hline
		U-Net-EffB2     &\xmark   & \xmark    &83.07   &91.73   &89.98  &83.23   &90.84\\
		U-Net-EffB2     &\xmark   & \cmark    &83.31   &\textbf{92.74}   &90.33  &84.37   &91.52\\
		U-Net-EffB2     &\cmark   & \cmark    &\textbf{84.09}   &92.23   &\textbf{91.57}  &\textbf{85.01}   &\textbf{91.90}\\
		\hline \hline
	\end{tabular}
\end{table*}

\subsection{Ensemble}
\label{Ensemble}
To boost segmentation performance, we use three distinct ensembling strategies, namely five-fold cross-validation, test time augmentation and result fusion.

\begin{figure}[b!]
\centering
\includegraphics[width=.95\linewidth]{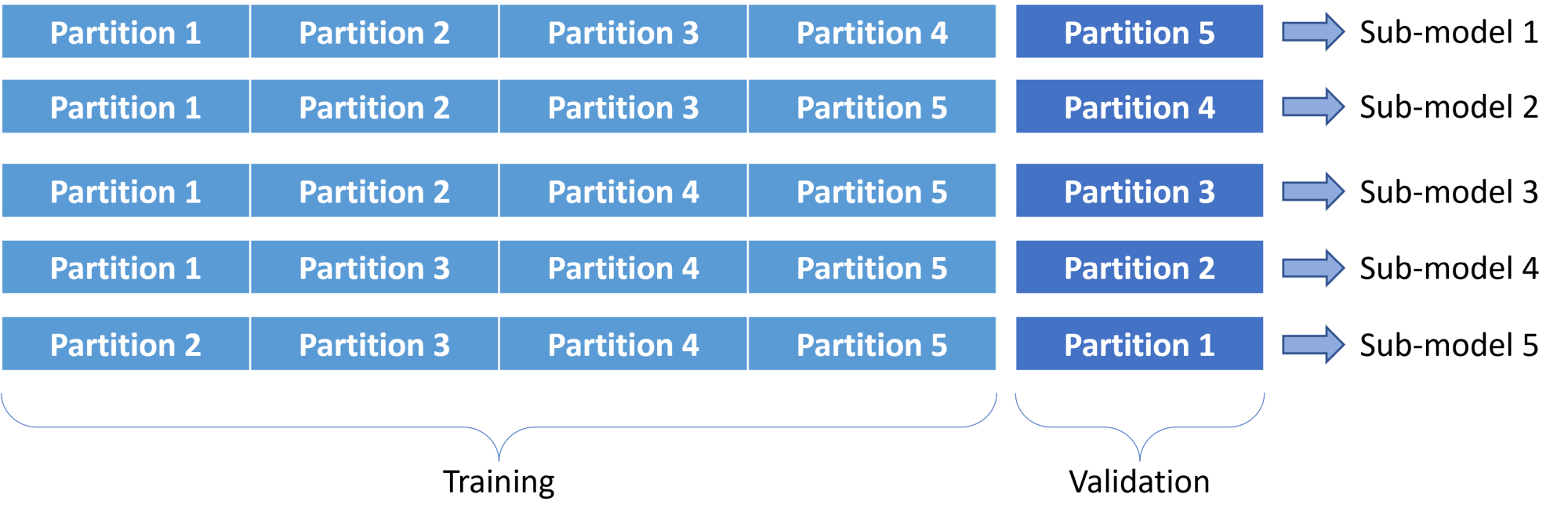}
\caption{Five-fold cross-validation training scheme. The entire training data is divided into 5 partitions. For each sub-model, 4 partitions are used for training, while the hold-out partition is used for validation. The generated sub-models are then used in the inference phase during ensembling.}
\label{5-fold}
\end{figure}

As illustrated in Fig.~\ref{5-fold}, instead of using the entire training set to train a single model, we divide the training data into 5 partitions and train one sub-model for each cross-validation split (i.e., for each of the sub-models, we use 4 partitions for training and the hold-out partition for validation). In the inference phase, we pass a test image to all 5 sub-models and take the average over the obtained results. We perform five-fold cross-validation for both U-Net and Link-Net models, and hence in total, 10 trained sub-models are generated.

As shown in former work on various image segmentation/classification tasks~\cite{pmlr-v156-bancher21a, 9412307}, test time augmentation (TTA) can boost the overall performance. We therefore employ TTA in the inference phase to yield improved segmentation. For this, we use 0, 90, 180, and 270-degree rotations as well as horizontal flipping. 

Last not least, since we train two distinct models -- a LinkNet with an EfficientNetB1 backbone and a U-Net with an EfficientNetB2 backbone -- we fuse their results in the inference phase. Various methods such as voting, bagging, or averaging exist for results fusion. We use straight-forward averaging to fuse the prediction probability masks from the two models. Averaging has been shown to yield comparable fusion performance to other more sophisticated ensembling methods for various medical image analysis tasks~\cite{9446924, pmlr-v156-bancher21a}.


\subsection{Post-processing}
\label{postprocess}
To form the final segmentation masks for a test image, we first binarise the fused prediction probability vectors using a 0.5 threshold. We further apply two post-processing steps, namely filling holes and removing very small detected objects, with the identical settings as described in~\cite{wang2020fully}. 

\subsection{Evaluation}
To evaluate the segmentation performance of our proposed method and also to compare the results to other state-of-the-art models, we use precision (Equ.~\ref{precision}), recall (Equ.~\ref{recall}), and data-based Dice score (Equ.~\ref{data_based_dice}) as suggested in~\cite{wang2020fully}. For the chronic wound dataset, we also report the data-based intersection over union (IoU) (Equ.~\ref{data_based_IOU}) score and the image-based Dice score (Equ.~\ref{image_based_dice}) for our proposed method. For the FUSeg challenge, we report the data-based Dice score, precision and recall as these scores were provided by the challenge organisers (the data-based Dice score was used as the main evaluation index in the challenge)~\cite{wang2022fuseg}. 

Based on the standard definitions of true positives ($TP$), false positives ($FP$), false negatives ($FN$), and with $N$ the total number of test images, precision is defined as 
\begin{equation}
\label{precision}
Precison = \frac{\sum_{i=1}^{N}  TP_i}{\sum_{i=1}^{N} TP_i + \sum_{i=1}^{N}  FP_i} ,
\end{equation}
recall as
\begin{equation}
\label{recall}
Recall = \frac{\sum_{i=1}^{N}  TP_i}{\sum_{i=1}^{N} TP_i + \sum_{i=1}^{N}  FN_i} ,
\end{equation}
the data-based Dice score as
\begin{equation}
\label{data_based_dice}
Dice_{data} = \frac{\sum_{i=1}^{N}  2TP_i}{\sum_{i=1}^{N} 2TP_i + \sum_{i=1}^{N}  FP_i + \sum_{i=1}^{N}  FN_i} ,
\end{equation}
the data-based IoU as
\begin{equation}
\label{data_based_IOU}
IoU_{data} = \frac{\sum_{i=1}^{N}  TP_i}{\sum_{i=1}^{N} TP_i + \sum_{i=1}^{N}  FP_i + \sum_{i=1}^{N}  FN_i} ,
\end{equation}
and the image-based Dice score as
\begin{equation}
\label{image_based_dice}
Dice_{image} = \frac{1}{N} \sum_{i=1}^{N}\frac{2TP_i}{2TP_i + FP_i + FN_i} .
\end{equation}

\begin{table*}[ht!]
	\caption{Top five performers of the MICCAI 2021 FUSeg challenge.}
	\centering
	\label{miccai}
	\begin{tabular}{lcccc}
		\hline \hline
		\textbf{team}  &\textbf{approach}  & \textbf{precision [\%]} & \textbf{recall [\%]}& \textbf{\begin{tabular}[c]{@{}l@{}}data-based\\ Dice [\%]\end{tabular}}\\
		\hline \hline
		Mahbod {\it et al.}  & this paper &  \textbf{91.55} & 86.22& \textbf{88.80} \\
		Zhang {\it et al.}  & U-Net with HarDNet68  &88.87 &\textbf{86.31} &87.57 \\
		Galdran {\it et al.}~\cite{galdran2021double} &  Stacked U-Nets &90.03 &84.00&86.91  \\
		Hong {\it et al.}  &  n/a  & n/a&n/a&86.27 \\
		Qayyum {\it et al.} &  U-Net with ASPP &n/a  &n/a&82.29 \\
		\hline \hline
	\end{tabular}
\end{table*}

\begin{figure}[b!]
	\centering
	\includegraphics[width=0.9\linewidth]{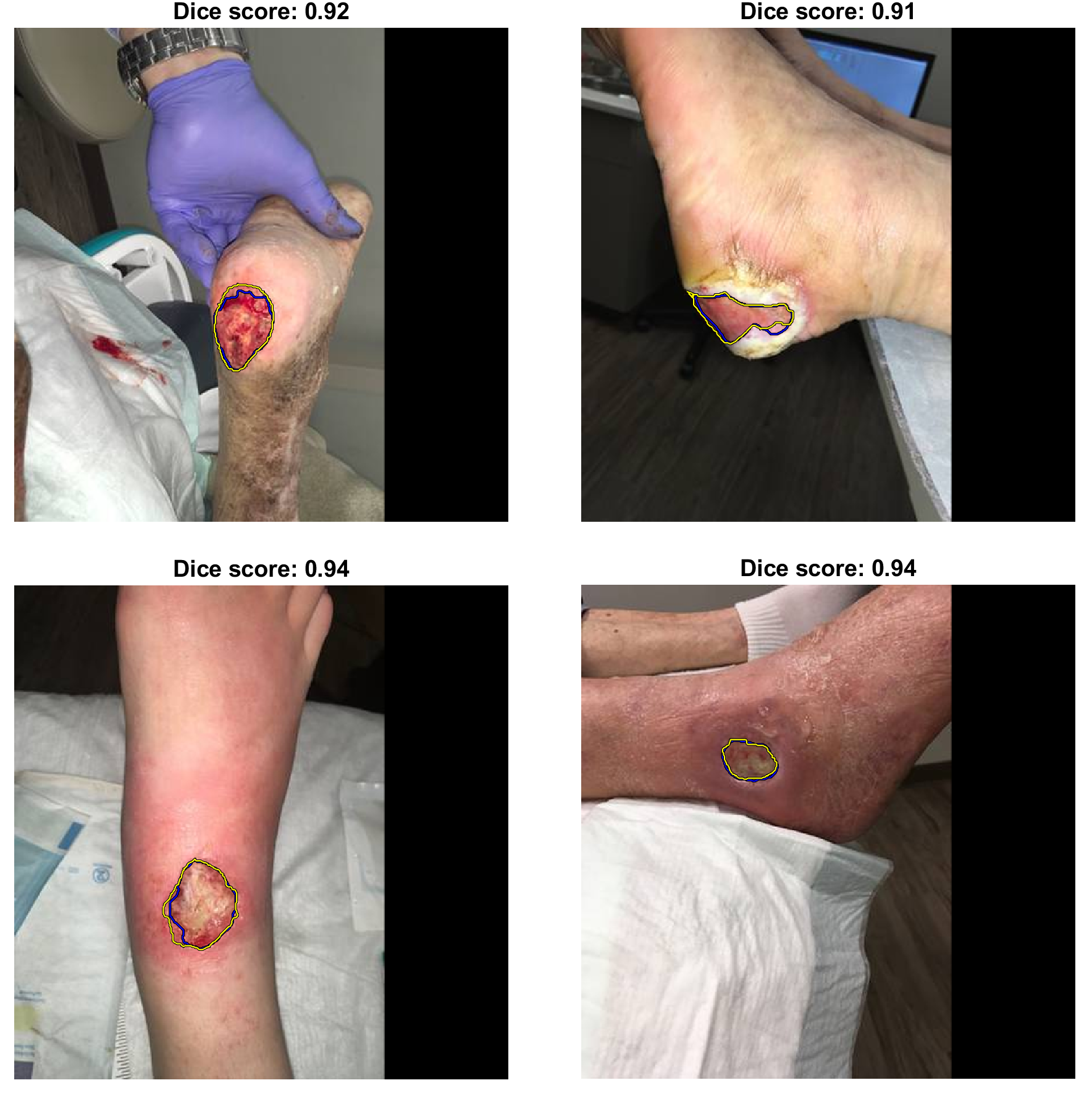}
	\caption{Examples from the chronic wound test set with acceptable segmentation performance. Yellow lines show the manual ground truth, and blue lines the predicted segmentation masks. }
	\label{good}
\end{figure}

\begin{figure}[b!]
	\centering
	\includegraphics[width=0.9\linewidth]{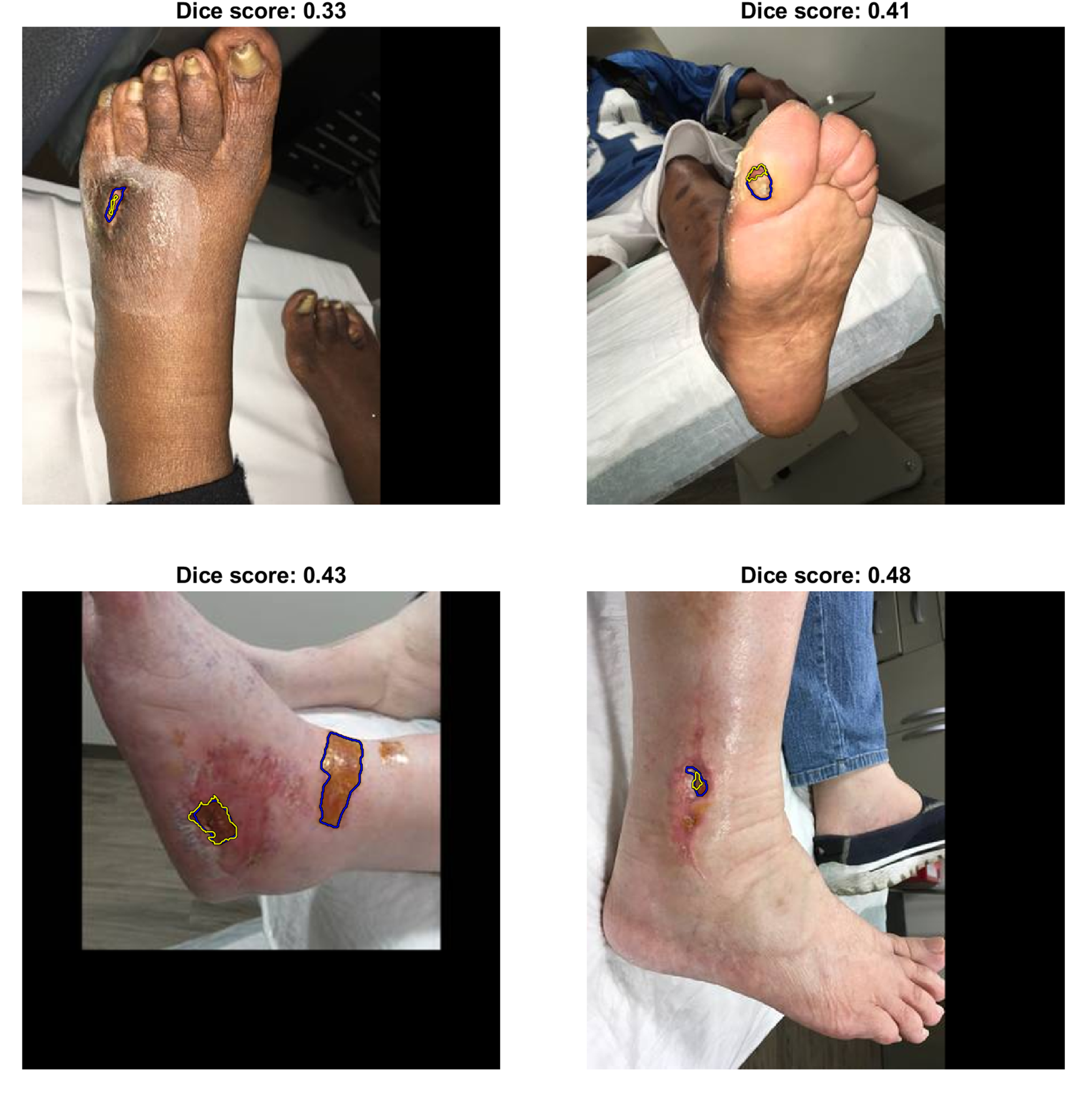}
	\caption{Examples from the chronic wound test set with poor segmentation performance (Dice scores below 60\%). The yellow lines show the manual ground truth in the images, and the blue lines show the predicted segmentation masks.}
	\label{poor-not-zero}
\end{figure}

\begin{figure}[t!]
	\centering
	\includegraphics[width=0.9\linewidth]{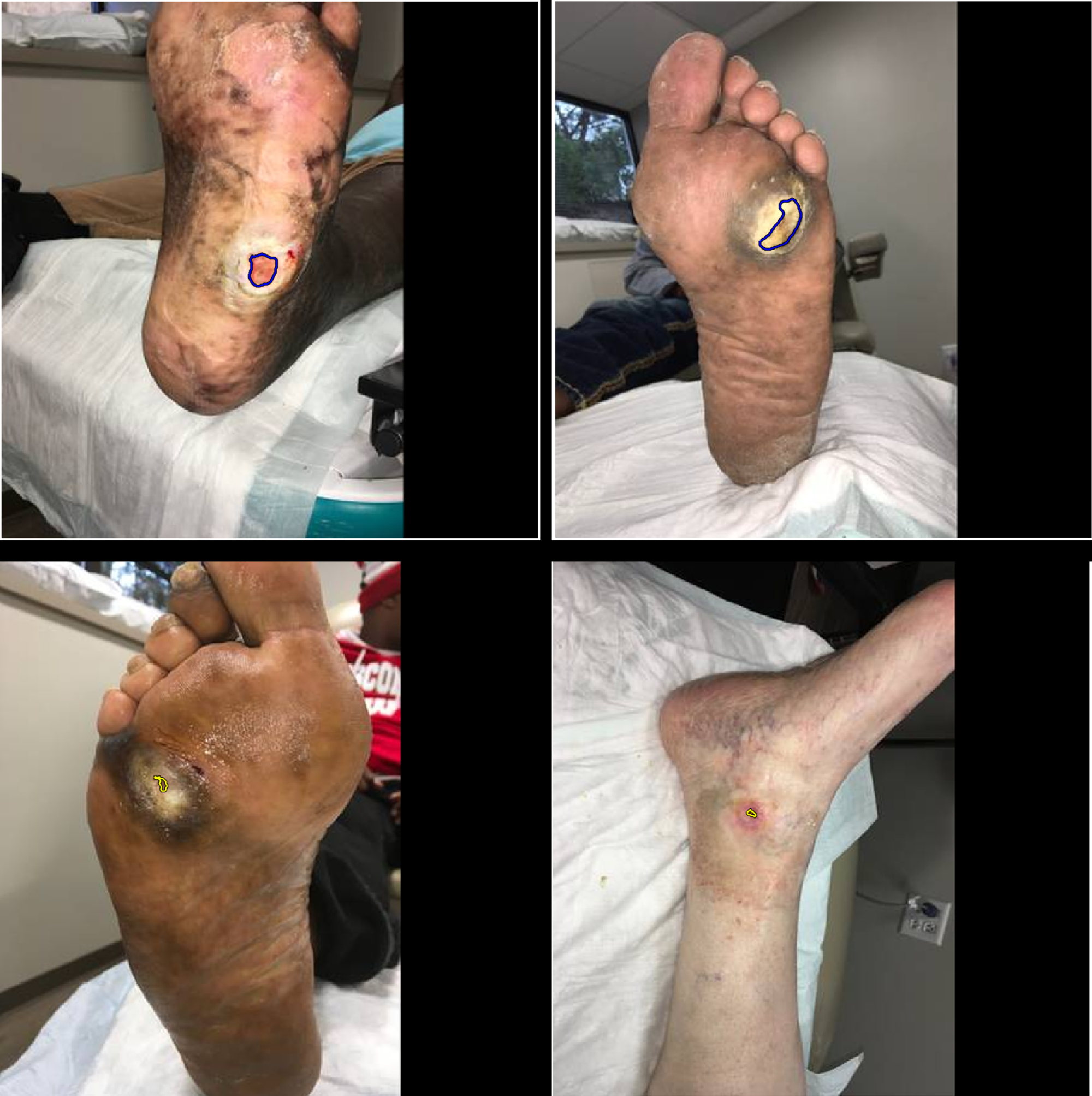}
	\caption{Examples from the chronic wound test set with a zero Dice score. The yellow lines show the manual ground truth in the images, and the blue lines show the predicted segmentation masks.} 
	\label{complex}
\end{figure}

\section{Results and Discussion}
\label{sec:results}
We report the results for the test set of the chronic wound dataset in Table~\ref{wound_res}. There, we also compare our proposed method with a number of deep learning-based approaches reported in~\cite{wang2020fully}. For our method, we give the results of the LinkNet model (with EfficientNetB1 backbone) and the U-Net model (with EfficientNetB2 backbone) on their own as well as the results of the fused model. 

As we can see, even on their own, the LinkNet and U-Net models outperform the other state-of-the-art methods for most evaluation indexes (all except precision). It is equally evident that the final fusion scheme yields another small improvement and thus excellent segmentation performance in comparison to the other state-of-the-art models. Moreover, comparing the U-Net results with the U-Net-EffB2 results shows the advantage of using a pre-trained CNN in the encoder part of the segmentation model. 

To show the effectiveness of the utilised ensemble strategies, we perform an ablation study and report the results in Table~\ref{ablation}. As the results for each of the models show, both the use of five-fold cross-validation and TTA results in better segmentation performance.

As can be inferred from Table~\ref{wound_res}, our proposed method delivers excellent segmentation performance for most of the images, and we show some examples with acceptable segmentation scores in Fig.~\ref{good}. However, for 18 of the 200 test images, relatively poor Dice scores below 60\% are obtained, where in 10 cases the segmentation performance is very poor with a zero Dice score. In these latter cases, there is either no lesion in the image but a false positive area is predicted (3 cases), or there are very small lesions in the images that are missed by the algorithm (6 cases), while the remaining image has an incorrect manual annotation. Some examples with poor (Dice score between 0\% and 60\%) and very poor segmentation results (Dice score of 0\%) are shown in Fig.~\ref{poor-not-zero} and Fig.~\ref{complex}, respectively.

With the presented methodology, we attended the MICCAI 2021 FUSeg challenge and submitted our inference codes and saved models in the frame of a Dockerised container. The results from our approach and the top four participating teams of the challenge are reported in Table~\ref{miccai}. These results were directly calculated by the challenge organisers and are available in the challenge leaderboard\footnote{\url{https://uwm-bigdata.github.io/wound-segmentation/}. As the challenge remains open for new post submissions, the ranking may change in future.}. As we can observe from Table~\ref{miccai}, our method outperforms the other approaches and is thus top ranked in the challenge based on the data-based Dice score. Our method also achieves the best precision performance and a very competitive recall score compared to the other approaches.

\section{Conclusions}
\label{conclusion}
Computer-aided foot ulcer segmentation can provide an efficient and effective alternative to manual analysis and subsequently calculated wound area measures. In this paper, we have proposed a powerful deep learning-based technique to segment foot ulcers in clinical images. Our approach incorporates two CNN segmentation networks and various ensembling strategies for improved segmentation performance. Results on two benchmark datasets demonstrate our method to yield excellent segmentation performance and to outperform other state-of-the-art deep learning-based algorithms, while being top-ranked in the recent MICCAI 2021 FUSeg challenge.

\section*{Acknowledgements}
This  project was  supported by  the  Austrian Research Promotion Agency (FFG), No. 872636. We would like to also thank NVIDIA for their generous GPU donation. This research study was conducted retrospectively using human subject data made available in open access form in former studies. Ethical approval was not required as confirmed by the license attached with the open access data.

\bibliographystyle{IEEEtran}
\bibliography{icpr22}

\end{document}